\documentclass[pra,aps,twocolumn,showpacs]{revtex4-1}     
\usepackage{epsfig,amsmath} 

\begin{document}

\title{Integer quantum Hall effect of interacting electrons in graphene} 

\author{Xin-Zhong Yan$^{1}$ and C. S. Ting$^2$}
\affiliation{$^{1}$Institute of Physics, Chinese Academy of Sciences, P.O. Box 603, 
Beijing 100190, China\\
$^{2}$Texas Center for Superconductivity, University of Houston, Houston, Texas 77204, USA}

\date{\today}

\begin{abstract}
By taking into account the charge and spin orderings and the exchange interactions between all the Landau levels, we investigate the integer quantum Hall effect of electrons in graphene using the mean-field theory. We find that the fourfold degeneracy of the Landau levels cannot be completely lifted by the Coulomb interactions. In particular, at fillings $\nu = 4n+2$ with $n = 0, 1, \cdots$, there is no splitting between the four-fold degenerated Landau levels. We show that with doping the degenerated lowest empty level can be sequentially filled one by one; the filled level is lower than the empty ones because of the Coulomb-exchange interactions. This result explains the step $\Delta\nu$ = 1 in the quantized Hall conductivity. We present a highly efficient method for dealing with huge number of the Coulomb couplings between all the Landau levels of the Dirac fermions.
\end{abstract}

\pacs{73.43.Cd,71.70.-d,73.22.Pr,72.80.Vp} 
 
\maketitle

\section{Introduction}

The study of quantum Hall effect (QHE) is an important aspect of graphene physics. By the noninteracting electron model, the Hall conductivity is given by $\sigma_{yx} = \nu e^2/h$ and $\nu = \pm (4n+2)$ with $n = 0, 1, \cdots$ as the index of highest-occupied Landau level (LL) as observed in the very early experiments \cite{Novoselov,Zhang1,Martin}. The step $\Delta\nu = 4$ stems from the spin and valley degeneracy of LLs of electrons in graphene. In addition to the fillings $\nu = \pm 2, \pm 6, \cdots$, the experiments then observed the states of $0, \pm 1$, and $\pm 4$ and fractional fillings at strong magnetic field \cite{Zhang,Jiang}. Later, all the integer and some fractional factors of $|\nu| \le 10$ in high-quality suspended graphene even at weak field ($<$1 T) \cite{Feldman} as well as $|\nu| \le 14$ in graphene on hexagonal boron nitride substrates at strong field \cite{Young,Yu,Amet,Chiappini} were observed. The appearance of these states was attributed to the $SU(4)$ symmetry breaking of electron system \cite{Lukose} including ferromagnetization (FM) \cite{Nomura}, FM with disorders \cite{Sheng}, the external magnetic field catalyzed canted antiferromagnetic spin ordering (ferromagnetic in the easy axis and antiferromagnetic in the easy plane) or charge-density wave \cite{Gusynin,Herbut,Kharitonov,Lado,Roy,Khveshchenko,Alicea,Gusynin1,Jung,Gorbar}, the field dependent Peierls distortion \cite{Fuchs}, and the Kekul\'e ordering \cite{Kharitonov}. The problem has been studied with models of short-range interactions \cite{Herbut,Kharitonov,Lado,Roy} and long-range Coulomb interactions \cite{Lukose,Nomura,Sheng,Khveshchenko,Alicea,Gusynin1,Jung,Gorbar}. Since the Coulomb interaction $V(q) = 2\pi e^2/q$ with $q$ as the momentum transfer between electrons scales as $1/\sqrt{B}$ (because of $q \propto \sqrt{B}$) under the Landau quantization in a magnetic field $B$, the long-range exchange interactions should be relevant to the QHE at weak field other than the short-range interactions (that are constants). For the clean system at weak field, the long-range Coulomb interactions should play the predominant role in determining the QHE of electrons. A realistic microscopic model should contain the long-range Coulomb interactions between electrons.

At weak magnetic field, there is a huge number of LLs in the valence and conduction bands. It has been a difficult task to deal with the Coulomb couplings between all these levels. So far, the long-range Coulomb interactions are treated within only a single level in most existing theories \cite{Nomura,Sheng,Alicea,Khveshchenko} or within very limited levels \cite{Gorbar}, which are valid at very strong magnetic field. To avoid manipulating the Coulomb couplings between the LLs, Ref. \onlinecite{Lukose} adopts the variational approach. Since how to treat the Coulomb couplings between all the Landau levels is a fundamental problem, it is necessary to develop a highly efficient method. 

In this paper, using the mean-field theory (MFT), we formulate the integer QHE (IQHE) of electrons in graphene taking into account the long-range Coulomb interactions as well as the on-site interaction. To overcome the numerical difficulty, we develop a highly efficient method for dealing with the Coulomb couplings between all the LLs. The QHE with $\Delta\nu = 1$ is usually considered as lifting of the fourfold degeneracy of the LLs. We will show that the fourfold degeneracy cannot be completely lifted for the electrons with Coulomb interactions. In particular, at $\nu = \pm (4n+2)$, the degeneracy is still 4. We will show that with doping the degenerated empty level can be sequentially filled one by one. For the interacting electrons, the LLs are not rigid but vary with the electron doping; the highest-occupied level is always lower than the lowest-unoccupied level, although they might be originally degenerated before the doping. 

\section{Formalism}

We begin with the description of the electron system in graphene. The honeycomb lattice of graphene shown in Fig. 1 (left) contains atoms $a$ and $b$ with lattice constant $a_0 \approx 2.46$ \AA. The Hamiltonian of the electrons with a neutralizing background is
\begin{equation}
H=-t\sum_{\langle ij\rangle s}c^{\dagger}_{is}c_{js}+U\sum_{j}\delta n_{j\uparrow}\delta n_{j\downarrow} +\frac{1}{2}\sum_{i\neq j}v_{ij}\delta n_{i}\delta n_{j} \nonumber\\
 \label{hmt}
\end{equation}
where $c^{\dagger}_{is}$ ($c_{is}$) creates (annihilates) an electron of spin $s$ in site $i$, $\langle ij\rangle$ means the sum over the nearest-neighbor (NN) sites, $t \approx$ 3 eV is the NN hopping energy \cite{Tatar,LMZ}, $\delta n_{is}=n_{is}-n_s$ is the number deviation of electrons of spin $s$ at site $i$ from the average occupation $n_s$, and $U$ and $v_{ij}$ are the Coulomb interactions between electrons. In real space, $v_{ij} = v(r_{ij})$ with $r_{ij}$ the distance between sites $i$ and $j$ is given by
\begin{equation}
v(r) = \frac{e^2}{r}[1-\exp(-q_0r)], \label{int}
\end{equation}
where $q_0$ is a parameter taking into account the effect of wave function spreading. Here we take $q_0 = 0.5/a_0$. 

\begin{figure}[t]
\centerline{\epsfig{file=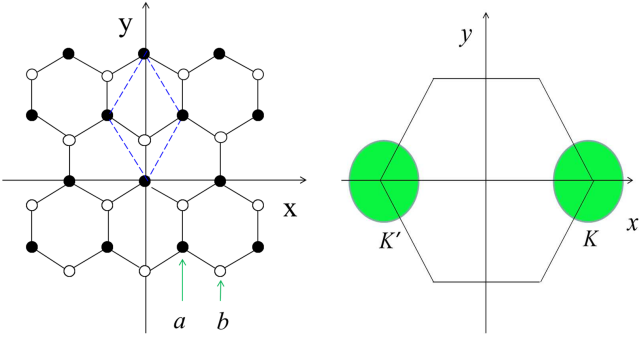,width=8.8 cm}}
\caption{(color online) Left: Lattice structure of graphene contains atoms $a$ (black) and $b$ (white). The dashed diamond is the unit cell.  Right: First Brillouin zone and the two valleys $K$ and $K'$ in the momentum space.} 
\end{figure} 

Here we treat the interactions between electrons by the MFT. In the Hartree term, there are spin and charge orderings with the order parameters defined as $m_j = (\langle\delta n_{j\uparrow}\rangle-\langle\delta n_{j\downarrow}\rangle)/2$ and $\rho_j = (\langle\delta n_{j\uparrow}\rangle+\langle\delta n_{j\downarrow}\rangle)$, respectively. These parameters depend only on the sublattice index $l$ (= $a$ or $b$), $m_j = m_l$ and $\rho_j = \rho_l$, where the position $j$ belongs to the sublattice $l$. There is only one parameter $\rho_a = -\rho_{b} \equiv \rho$ for the charge ordering because of the charge neutrality. For the Fock term, we take the screening effect (due to the charge-density fluctuations) \cite{Luttinger} in the exchange interaction by the Thomas-Fermi (TF) screening function $\epsilon (q) = 1 + q_{TF}/q$ with $q_{TF}$ as the TF wave number and $q$ the momentum transfer between electrons. By translating the lattice to the continuous space, the Hamiltonian under the MFT is obtained as (see Appendix A)
\begin{eqnarray}
H&=&\sum_{vs}[\int d\vec{r}C^{\dagger}_{vs}(r)h_{v}(\vec p)C_{vs}(r) \nonumber\\
&&+\int d\vec{r}\int d\vec{r'}C^{\dagger}_{vs}(r)\Sigma^{vs}(r,r')C_{vs}(r')] \label{mfh}
\end{eqnarray}
where $h_{v}(\vec p) = v_0(s_vp_x\sigma_1+p_y\sigma_2)$ with $\vec p$ the momentum operator and $v_0 = \sqrt{3}ta_0/2\hbar$ as the Fermi velocity, $s_v = 1$ $(-1)$ for electrons in valley $v = K$ $(K'=-K)$ [see Fig.1 (right)], the Pauli matrices $\sigma$'s operate in the space ($a$, $b$) of sublattices, $C^{\dagger}_{vs}(r) = [c^{\dagger}_{avs}(r),c^{\dagger}_{bvs}(r)]$ with $c^{\dagger}_{a(b)vs}(r)$ creating an electron of spin $s$ and valley $v$ on position $r$ in the $a$ ($b$) sublattice, and $\Sigma^{vs}(r,r')$ is a 2$\times$2 matrix of the self-energy. We will use the units in which $\hbar = a_0 = v_0$ = 1. The energy unit then is $\epsilon_0 = \hbar v_0/a_0$ = 1.

Under a magnetic field $B$ perpendicular to the graphene plane, we take the vector potential as $\vec A(r) = (0,Bx)$. The momentum along the $y$-direction is a good number and is denoted as $k$. For going to the LL picture, we expand the operator $C_{vs}(r)$ as
\begin{equation}
C_{vs}(r)=\sum_{nk\lambda}\phi^v_{nk}(r)\psi^{vs}_{\lambda}(n)\hat a^{vs}_{\lambda nk} \label{cr}
\end{equation}
where $\phi^v_{nk}(r)$ is a $2\times$2 matrix given by 
\begin{eqnarray}
\phi^{v=K}_{nk}(r) &=& \frac{e^{iky}}{\sqrt{L}}\begin{pmatrix}
\phi_{n-1}(x-x_c)&0\\
0&i\phi_{n}(x-x_c)\\
\end{pmatrix},\nonumber\\
\phi^{v=K'}_{nk}(r) &=& \frac{e^{iky}}{\sqrt{L}}\begin{pmatrix}
\phi_{n}(x-x_c)&0\\
0&i\phi_{n-1}(x-x_c)\\
\end{pmatrix} \label{wv}
\end{eqnarray}
with $L$ the length of the lattice along the $y$ direction and $\phi_n(x-x_c)$ the $n$th level wave function of a harmonic oscillator along the $x$ direction with the center at $x_c = -k/B$, $\psi^{vs}_{\lambda}(n)$ (real) is a two-component spinner, and $\hat a^{vs}_{\lambda nk}$ annihilates an electron of momentum $k$ (along the $y$ direction) and spin $s$ in valley $v$ at $\lambda$th LL of index $n$. For $n < 0$, $\phi_n$ is understood as 0. In the LL picture, the elements of the self-energy matrix are given by (see Appendix A)
\begin{equation}
\Sigma^{vs}_{ll'}(n)=(v_c\rho_l-sUm_l)\delta_{ll'}-\sum_{n'}v^v_{ll'}(n,n')g^{vs}_{ll'}(n'), \label{sf}
\end{equation}
with $v_c$ a potential for charge ordering and $s$ = 1 (-1) for spin up (down), and the two matrices $v^v(n,n')$ and $g^{vs}(n')$ are given by
\begin{eqnarray}
v^{K}(n,n')&=&\sigma_1v^{K'}(n,n')\sigma_1\nonumber\\
&=&\int_0^{\infty}\frac{qdq}{2\pi}v^{sc}(q)e^{-\xi}\xi^mJ_{n,n'}(\xi)\otimes J^{t}_{n,n'}(\xi)\nonumber\\
\label{vkn}\\
J^{t}_{n,n'}(\xi) &=& [\sqrt{\frac{(n_2-1)!}{(n_1-1)!}}L^m_{n_2-1}(\xi), \sqrt{\frac{n_2!}{n_1!}}L^m_{n_2}(\xi)] \nonumber\\
g^{vs}(n)&=&\sum_{\lambda}[f^{vs}_{\lambda}(n) -1/2]\psi^{vs}_{\lambda}(n)\otimes \psi^{vs,t}_{\lambda}(n) \nonumber
\end{eqnarray}
where $v^{sc}(q)=v(q)/\epsilon(q)$ is the screened interaction, $\xi=q^2/2B$, $J^{t}_{n,n'}$ ($\psi^{vs,t}_{\lambda}$) is a transpose of $J_{n,n'}$ ($\psi^{vs}_{\lambda}$), $L^m_{n_2}(\xi)$ is the associated Laguerre polynomial, $n_1 = \max(n,n')$, $n_2 = \min(n,n')$, $m = |n-n'|$, and $f^{vs}_{\lambda}(n)$ is the Fermi distribution of valley-$v$ and spin-$s$ electrons in the $\lambda$th level of index $n$. 
The wave number $q_{TF}$ in the TF screening function $\epsilon (q) = 1 + q_{TF}/q$ is given by $q_{TF} = 8\sqrt{\pi\delta/\sqrt{3}}e^2/a_0v_0$ where $\delta$ is the doped electron number per atom.

The LLs $E^{vs}_{\lambda n}$ are determined by
\begin{equation}
[\sqrt{2Bn}\sigma_1+\Sigma^{vs}(n)]\psi^{vs}_{\lambda}(n)=E^{vs}_{\lambda n}\psi^{vs}_{\lambda}(n). \label{ll}
\end{equation}
Express the self-energy matrix as $\Sigma^{vs}(n) = \Sigma^{vs}_0(n)\sigma_0+\Sigma^{vs}_1(n)\sigma_1+\Sigma^{vs}_3(n)\sigma_3$. The energy levels for $n\ne 0$ are obtained as
\begin{eqnarray}
E^{vs}_{\lambda n}&=&\Sigma^{vs}_0(n)+s_{\lambda}\{[\sqrt{2Bn}+\Sigma^{vs}_1(n)]^2+[\Sigma^{vs}_3(n)]^2\}^{1/2} \nonumber\\
&\equiv&\Sigma^{vs}_0(n)+s_{\lambda}E^{vs}(n), ~~~~\lambda = +,-. \label{engy}
\end{eqnarray}
The wavefunctions are
\begin{eqnarray}
\psi^{vs}_{+}(n) &=& 
\left[\begin{array}{c}R^{vs}_+(n)\\
R^{vs}_-(n)
\end{array}\right],\nonumber\\
\psi^{vs}_{-}(n) &=& 
\left[\begin{array}{c}-R^{vs}_-(n)\\
R^{vs}_+(n)
\end{array}\right] \label{wvf}
\end{eqnarray}
where $R^{vs}_{\pm}(n) = \sqrt{1\pm\Sigma^{vs}_3(n)/E^{vs}(n)}/\sqrt{2}$. For $n = 0$, the eigenstates are given by
\begin{eqnarray}
E^{Ks}_0&=&\Sigma^{vs}_{bb}(0), ~~~~ \psi^{Ks}(0) = 
\left[\begin{array}{c}0\\
1\end{array}\right], \nonumber\\
E^{K's}_0&=&\Sigma^{vs}_{aa}(0), ~~~~ \psi^{K's}(0) = 
\left[\begin{array}{c}1\\
0\end{array}\right]. 
\end{eqnarray}
The charge and spin orders are calculated by
\begin{eqnarray}
\rho&=&\frac{s_0B}{4\pi}\sum_{l\lambda n vs}s_lf^{vs}_{\lambda}(n)|\psi^{vs}_{l\lambda}(n)|^2 \label{rho}\\
m_l&=&\frac{s_0B}{4\pi}\sum_{\lambda n vs}sf^{vs}_{\lambda}(n)|\psi^{vs}_{l\lambda}(n)|^2 \label{ml}
\end{eqnarray}
where $s_0 = \sqrt{3}/2$ is the area of the unit cell, $B/2\pi$ is the spatial degeneracy of the Landau state, $\psi^{vs}_{l\lambda}(n)$ is the $l$th component of $\psi^{vs}_{\lambda}(n)$ and $s_l$ = 1 (-1) for $l = a$ ($b$). By taking the Hubbard $U/\epsilon_0 = 2.5$ \cite{Gloora}, the potential $v_c$ is then determined as $v_c/\epsilon_0 = 0.173$ (see Appendix A).

\begin{figure}
\centerline{\epsfig{file=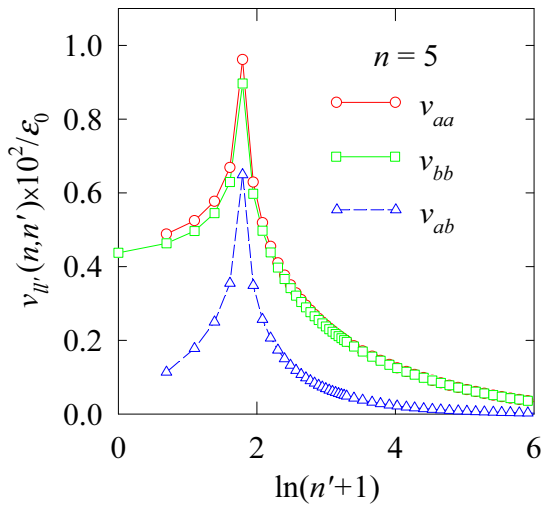,width=7.5 cm}}
\caption{(color online) $K$-valley interaction $v_{ll'}(n,n')$'s as functions of $n'$ at $n = 5$ and $B = 0.5$ T and $\nu$ = 0. $v_{aa}(n,n')$ and $v_{ab}(n,n')$ are defined for $n' \ge 1$.} 
\end{figure} 

Here, the magnetic field is in unit of $B_0 = \hbar c/ea^2_0 = 1.105\times 10^4$ T. Corresponding to the momentum cutoff $p_c \sim 1$, the largest Landau index is $N \sim 0.5/B$. For $B = 0.5$ T $= 0.45\times 10^{-4}B_0$, we have $N \sim 1.1\times 10^4$. According to Eq. (\ref{sf}), the number of the $2\times 2$ interaction matrix $v^v(n,n') = v^v(n',n)$ in the calculation is about $N^2/2$. There are two difficulties in the numerical calculation of the matrix element $v^v_{ll'}(n,n')$. First, $v^v_{ll'}(n,n')$ as given by Eq. (\ref{vkn}) for large $\min(n,n')$ cannot be precisely evaluated by the integral with the Laguerre polynomials involved. This problem is solved in Appendix B. Second, since the number of matrix elements is  of the order $N^2$, the requirement for their storage may exceed the memory limit of a computer. Even though there is no storage problem, it is still formidable work to self-consistently solve Eqs. (\ref{sf})-(\ref{ml}) because of the $N$ term summations at weak $B$. Most of the works study the case of strong magnetic field so the calculation is simplified by taking only one term of $n' = n$ in the sum in Eq. (\ref{sf}) \cite{Nomura,Sheng,Alicea,Khveshchenko}. In Fig. 2, $v_{ll'}(n,n')\equiv v^K_{ll'}(n,n')$'s are shown as functions of $n'$ at $n = 5$ and $B = 0.5$ T and $\nu = 0$. As seen from Fig. 2, the interactions $v^K_{ll'}(n,n')$ vary slowly with $n'$ at a given $n$. Therefore, the calculation taking only the peak-value term in the sum is not sufficient for reflecting the interaction effect. For overcoming this difficulty, we develop a highly efficient method in Appendix C. 

For completeness, we derive the quantum Hall conductivity in Appendix D.

\section{Numerical results}

We have self-consistently solved Eqs. (\ref{sf})-(\ref{ml}). Shown in Fig. 3 are the self-energy $\Sigma_{ll'}(n) \equiv \Sigma^{K\uparrow}_{ll'}(n)$ at $B = 0.5$ T and $\nu$ = 0. In valley $K$, the self-energy elements $\Sigma^{K\uparrow}_{aa}(n)$ and $\Sigma^{K\uparrow}_{ab}(n)$ are defined for $n \geq$ 1. The element $\Sigma^{K\uparrow}_{ab}(n)$ describes the interaction exchange effect between the sublattices $a$ and $b$. Because of the induction by the inter-sublattice hopping, $\Sigma^{K\uparrow}_{ab}(n)$ is stronger than $\Sigma^{K\uparrow}_{aa}(n)$ and $\Sigma^{K\uparrow}_{bb}(n)$ for $n >$ 1. The diagonal parts $\Sigma^{K\uparrow}_{aa}(n)$ and $\Sigma^{K\uparrow}_{bb}(n)$ vary slowly with the index $n$. The diagonal parts with opposite signs mean a dynamical mass gap. 

\begin{figure}[t]
\centerline{\epsfig{file=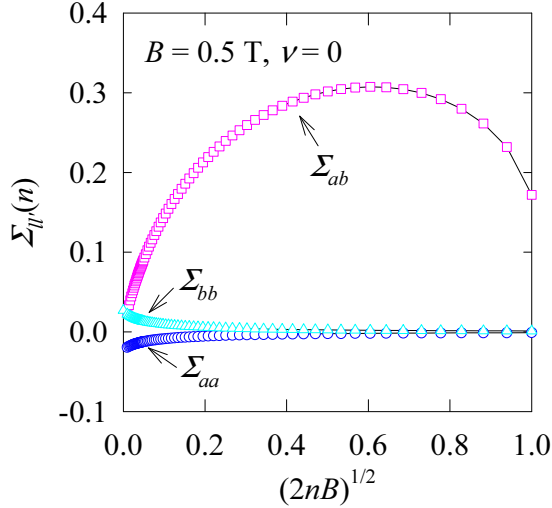,width=7.5 cm}}
\caption{(color online) Self-energy $\Sigma_{ll'}(n)$'s of spin-up electrons in valley $K$ as functions of $n$ at $B$ = 0.5 T and $\nu$ = 0. $\Sigma_{aa}(n)$ and $\Sigma_{ab}(n)$ are defined for $n \ge 1$.} 
\end{figure} 

The LLs at filling numbers $\nu$ = 0 to 6 are depicted in Fig. 4 for $B = 0.5$ T. At $\nu = 0$, the four-fold degeneracy of the levels of $n$ = 0 is partially lifted with a gap between the filled level of spin-up (down) electrons at the $K'$ ($K$)-valley and the empty level at the $K$ ($K'$) valley. The lowest empty level is still two-fold degenerated. According to the wave functions of $n = 0$ given by Eq. (\ref{wv}), the spin-up electrons occupy sublattice $a$ while the spin-down electrons occupy sublattice $b$. Such an occupation gives rise to the antiferromagnetic spin ordering (see Fig. 5). Since there is no charge ordering, $\rho =0$, the self-energy satisfies the relation $\sigma_1\Sigma^{Ks}(n)\sigma_1 = \Sigma^{K'\bar s}(n)$ with $\bar s = -s$, which means each level still has a degeneracy 2. At the charge neutrality point CNP, the levels satisfy the particle-hole symmetry [$\sigma_2\Sigma^{Ks}(n)\sigma_2 = -\Sigma^{K's}(n)$]: for a positive $K$-valley level, there is a negative $K'$-valley level of the same spin electrons. 

The question now is, with doping electrons, should the two empty levels of $n$ = 0 be filled or can there exist a state in which only one of them is filled? The former will give a quantum Hall state of $\nu =$ 2 and the latter will be $\nu = 1$. Our result shows that there can exist a state with the $K$ (or $K'$) valley level of $n = 0$ filled with spin-up (down) electrons [occupying sublattice $b$ ($a$)]. For the case of $\nu =1$ shown in Fig. 4, the energy of the top occupied level is $\Sigma^{K\uparrow}_{bb}(0) = v_c\rho_b-Um_b+ \Sigma^{K\uparrow,xc}_{bb}(0)$ where the last term is the exchange part of the self-energy. The empty level is $\Sigma^{K'\downarrow}_{aa}(0) = v_c\rho_a+Um_a+ \Sigma^{K'\downarrow,xc}_{bb}(0)$. In this state, above the spin ordering there is another symmetry breaking from the charge ordering. Since the sublattice $b$ now is occupied with more electrons than the sublattice $a$, the charge order parameter is obtained as $\rho = \rho_a = -\rho_b <$ 0 as shown in Fig. 5. Meanwhile, the spin ordering in sublattice $b$ is weakened with magnitude $|m_b|$ smaller than $|m_a|$. The charge ordering gives rise to a negative contribution to the gap $\Sigma^{K'\downarrow}_{aa}(0)-\Sigma^{K\uparrow}_{bb}(0)$ which means the Coulomb interactions do not favor the charge ordering. However, since the spin ordering and the exchange effect are strong enough in this state, the energy cost of the charge ordering is fully compensated. As a result, the imbalanced electron distributions in the two valleys and in two spins lead to the filled level being lower than the empty one. 

Here we go a step further to analyze the LLs changes due to the level $E^{K\uparrow}_0$ being filled. As seen from Fig. 4, with this level filled, all the levels shift toward zero energy. The self-energy change of ($K\uparrow$) electrons is 
\begin{eqnarray}
\Delta\Sigma^{vs}_{ll'}(n)|_{vs=K\uparrow}&\approx&(v_c\Delta\rho_l-U\Delta m_l)\delta_{ll'}\nonumber\\
 && -v^K_{ll'}(n,0)\psi^{vs}_{l\lambda}(0)\psi^{vs}_{l'\lambda}(0)|_{vs=K\uparrow,\lambda=+} \nonumber\\
&=&(v_c\Delta\rho_l-U\Delta m_l)\delta_{ll'}-v^K_{ll'}(n,0)\delta_{ll'}\delta_{lb} \nonumber\\
\label{sft}
\end{eqnarray}
temporarily neglecting the changes due to the wavefunction changes of filled levels in the valence band. For $n = 0$, the energy level shift $\Delta E^{K\uparrow}(0) < 0$ comes mainly from the last term in Eq. (\ref{sft})\-the long-range Coulomb exchange. For other states, the effects of $\Delta\rho_l$ and $\Delta m_l$ and the wave function changes of the filled levels are equally important to the self-energy and thereby the LLs. Among these levels, the largest change is the level $E^{K\uparrow}(0)$. The analysis here is also valid for other fillings. In short, when a level is filled with doping, the level not only shifts itself, but also influences other levels through the Coulomb interactions.

With further doping, the only empty level in valley $K'$ for spin-down of $n = 0$ will be filled, giving rise to the state of $\nu = 2$.

\begin{figure}[t]
\centerline{\epsfig{file=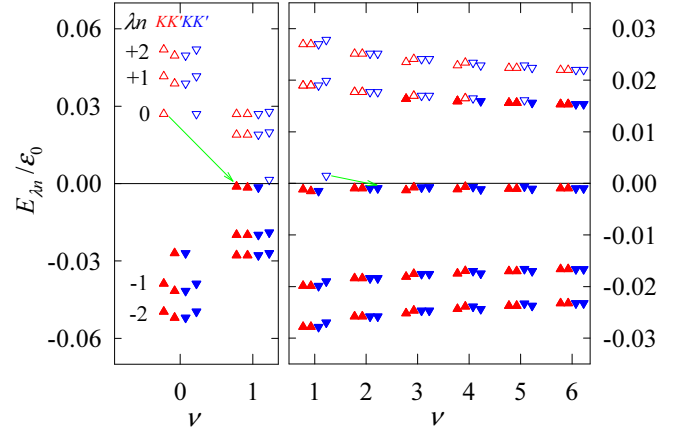,width=8.8 cm}}
\caption{(color online) Landau levels $E_{\lambda n}$ at $B = 0.5$ T. The red-filled (empty) upward triangles for occupied (empty) levels of spin-up electrons in valleys $K$ and $K'$. The blue downward-triangles are for the corresponding levels of spin-down electrons. In each column corresponding to a spin and valley at filling number $\nu$, the five levels are for $n = 0$ and $\pm n$ ($n = 1, 2$) with $+n$ ($-n$) standing forthe upper (lower) energy level of index $n$. An arrow indicates the energy change of a level from an empty to a filled state.} 
\end{figure} 

In the state of $\nu = 2$, the orderings vanish $\rho_l = m_l = 0$ and the self-energy satisfies the symmetry $\sigma_1\Sigma^{vs}(n)\sigma_1 = \Sigma^{\bar v s}(n)$ with $\bar v = -v$ and becomes independent of $s$. Therefore, the state of $\nu$ = 2 is a high symmetry state with degeneracy 4. We again go back to a similar question as mentioned above, with further doping electrons, should the four empty levels of $n = 1$ all be filled or can there exist a state in which only one of them is filled?
The answer is the four levels can be sequentially filled one by one. Although they are degenerated at $\nu = 2$, the filled levels are lower than the empty levels which can be seen from Fig. 4. For $\nu > 2$, since the Coulomb interactions are screened, the orderings are weak. For $\nu = 3$ and 5, since more levels are occupied with the spin-up electrons, we get the ferromagnetic states with unequal magnitudes of spin orderings in the two sublattices as shown in Fig. 5. Corresponding to the imbalanced occupation in the two valleys and thereby in the two sublattices, the charge ordering is finite at $\nu = 3$ and 5. At $\nu = 4$ where the levels for the spin-up electron in valley $K$ and the spin-down electron in valley $K'$ are filled, the symmetry $\sigma_1\Sigma^{vs}(n)\sigma_1 = \Sigma^{\bar v\bar s}(n)$ leads to a degeneracy 2. Because the numbers of electrons occupied in the two sublattice are equal, the charge ordering vanishes. But the spins in the two sublattices are not balanced; the system is in the antiferromagnetic state. At $\nu = 6$, all of the upper four levels of $n = 1$ are filled and we again reach a high symmetric state with $\rho = m_l = 0$ as in $\nu = 2$. 

At higher filling numbers $\nu > 6$, the level filling processes are similar as that from $\nu = 2$ to 6. But the gap between the highest filled level and the lowest empty level of an index $n$ decreases with $n$ because the Coulomb coupling is less important at high $n$. Therefore, above certain large $\nu$, the step $\Delta\nu = 4$ in the Hall conductivity will be observed.

\begin{figure}[t]
\centerline{\epsfig{file=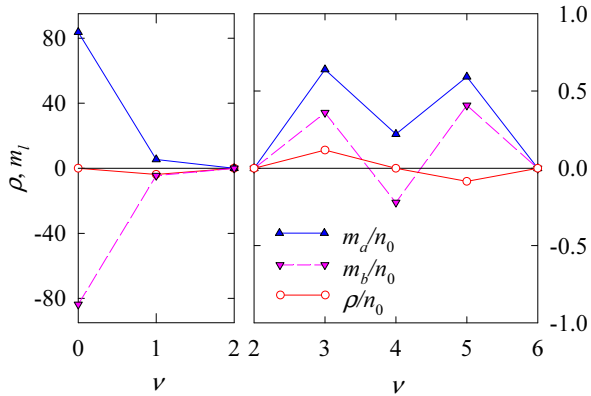,width=8. cm}}
\caption{(color online) Spin and charge ordering parameters as functions of filling number $\nu$. Here, $\rho \equiv \rho_a = -\rho_b$, and $n_0 = \sqrt{3}B/8\pi$ is the doped electrons per level per atom.} 
\end{figure} 

We have checked that all the orderings vanish at zero magnetic field at the CNP. The reason is that the density of states of the Dirac fermions vanishes at the zero energy. The spin ordering at finite $B$ at $\nu = 0$ occurs just because of the feature of the $n = 0$ Landau state. This original spin ordering has an effect on the symmetry breaking at subsequent fillings. In this sense, the spin and charge orderings are catalyzed by the external magnetic field. This result is consistent with the existing theories \cite{Gusynin,Herbut,Kharitonov,Lado,Roy,Khveshchenko,Alicea,Gusynin1,Jung,Gorbar}. 

The physics of sequentially filling a degenerated level is consistent with some of the existing works \cite{Lukose,Gorbar}.

It should be indicated that a constant term reflecting the gate voltage should be included within the square brackets in Eq. (\ref{ll}) for determining the absolute values of the levels. This term is given by $e^2n_0\nu/C\epsilon_0$ where $-e$ is the charge of an electron, $n_0 = \sqrt{3}B/8\pi$ is the number of doped electrons per atom per level, and $C$ is the charge capacity per atom. The magnitude of $C$ is dependent on the real system and is smaller than (or the same order) $2\times 10^{-5}e$/V of the system the graphene placed on a SiO$_2$ substrate \cite{Novoselov}. This term increases with $\nu$ much stronger than the decreasing of the levels shown in Fig. 4. As a result, the absolute values of the levels increase with the electron doping.  

We have neglected the Zeeman splitting. At $B = 0.5$ T, the splitting $\mu_B B/\epsilon_0 = 1.1\times 10^{-5}$ is much less than the smallest gap $\sim 5\times 10^{-4} \epsilon_0$ (= 1.3 meV) appearing at $\nu = 5$ shown in Fig. 4.  

Since the system satisfies the particle-hole symmetry \cite{Yan}, the above results for the electron doping can be converted to the case of hole doping.

\section{conclusion}

In conclusion, we have investigated the IQHE of electrons in graphene by the MFT taking into account the Coulomb couplings between all the Landau levels. At the fillings $\nu = 4n+2$, there are no spin and charge orderings and the Landau levels are four-fold degenerated because of the invariance of the system under the exchanges of spin and valley. We have shown that the lowest degenerated empty levels can be sequentially filled one by one with the filled levels lower than the empty levels. This filling process can exist because with doping the inter-electronic interactions give rise to the spin and valley dependent self-energy and the symmetry is broken. This filling process results in a step of $\Delta\nu = 1$ in the quantized Hall conductivity. 

We have developed a highly efficient method for dealing with a huge number of Coulomb couplings between all the Landau levels. With this method, we are able to study the QHS of interacting Dirac fermions at weak magnetic field.

\acknowledgments

This work was supported by the National Basic Research 973 Program of China under Grant No. 2016YFA0202300 and the Robert A. Welch Foundation under Grant No. E-1146.

\appendix
\section{Hartree-Fock approximation} 
\renewcommand{\theequation}{\thesection\arabic{equation}}
\setcounter{equation}{0}

By the MFT, the interactions in Eq. (\ref{hmt}) are approximated by the Hartree-Fock factorization. The direct part under the Hartree factorization reads
\begin{equation}
H_{dir} = \sum_{ljs}(-sUm_l+v_c\rho_l) c^{\dagger}_{ljs}c_{ljs}, \label{ht}
\end{equation}
where the sum runs over the sublattices $l$ and the sites $j$ of sublattice $l$ and spin $s, s = 1 (-1)$ for spin up (down), and $v_c$ is given by
\begin{equation}
v_c = U/2-v(r_0) + \sum_{\vec r\ne 0}[v(r)-v(|\vec r+\vec r_0|)], \label{vc}
\end{equation}
with $\vec r_0$ the vector from atom $a$ to atom $b$ in the unit cell and the $\vec r$-sum runs over the sites of sublattice $a$. For $U/\epsilon_0 = 2.5$, $v_c/\epsilon_0 = 0.173$ is obtained.

For the Fock term, the interactions between electrons include the screening effect due to the charge-density fluctuations \cite{Luttinger}. The exchange part under the Fock factorization is given by
\begin{equation}
H_{xc} = \frac{1}{2}\sum_{li\ne l'j,s}v^{sc}_{li,l'j}\langle[c_{lis}, c^{\dagger}_{l'js}]\rangle c^{\dagger}_{lis}c_{l'js}, \label{xc}
\end{equation}
where $v^{sc}_{li,l'j}$ is the screened interaction between electrons on sites $i$ of sublattice $l$ and $j$ of sublattice $l'$. Since the low-energy states of electrons close to the Dirac points are under consideration, the electron operator $c_{ljs}$ can be expanded as 
\begin{equation}
c_{ljs} = c_{ljKs}\exp(i\vec K\cdot\vec r_{lj})+c_{ljK's}\exp(-i\vec K\cdot\vec r_{lj}) \label{op}
\end{equation}
where $\vec K = -\vec K' = (4\pi/3,0), \vec r_{jl}$ is the position vector of lattice point $j$ on sublattice $l, \exp(\pm i\vec K\cdot\vec r_{lj})$ is the rapid phase factor, and $c_{ljK(K')s}$ is an operator slowly varying with the position. To rewrite Eq. (\ref{xc}) in terms of these new operators with valley index, by noting that the average of two operators of different valleys vanishes, we obtain
\begin{eqnarray}
H_{xc} &=& \frac{1}{2}\sum_{li\ne l'jvs}v^{sc}_{li,l'j}\{\langle[c_{livs}, c^{\dagger}_{l'jvs}]\rangle 
\nonumber\\
&&+
\langle[c_{li\bar vs}, c^{\dagger}_{l'j\bar vs}]\rangle\exp[-i2\vec K_v\cdot(\vec r_{li}-\vec r_{l'j})]
 \}c^{\dagger}_{livs}c_{l'jvs} \nonumber\\
&\approx& \frac{1}{2}\sum_{lil'j,vs}\{v^{sc}_{li,l'j}\langle[c_{livs}, c^{\dagger}_{l'jvs}]\rangle |_{li\ne l'j}\nonumber\\
&&~+v^{vJ}_{ll'}
\langle[c_{li\bar vs}, c^{\dagger}_{l'j\bar vs}]\rangle\delta_{ij}
 \}c^{\dagger}_{livs}c_{l'jvs} \label{xc1}
\end{eqnarray}
where $\bar v = \bar K (\bar K'$) = -$K (K), \vec K_K vec K_{K'}) = \vec K (-\vec K)$, the use of the property of $c^{\dagger}_{ljvs}$ slowly varying with position has been made in obtaining the last line, and $v^{vJ}_{ll'}$ is given by
\begin{eqnarray}
v^{vJ}_{ll'} = \sum_{j}v^{sc}_{li,l'j}\exp[-i2\vec K_v\cdot(\vec r_{li}-\vec r_{l'j})]|_{li\ne l'j}.
\nonumber 
\end{eqnarray}
The last term in Eq. (\ref{xc1}) contributes to the current ordering \cite{Alicea,Yan1} and gives rise to slight renormalization of the interactions $U$ and $v_c$. Since $v^{vJ}_{ll'}$'s are short-range interactions, as argued in the Introduction, they are less important than the long-range Coulomb interactions in the present case. Here we neglect the contribution from the last term in Eq. (\ref{xc1}) for simplifying the equations. 

When going to the continuous space, we take the following replacement
\begin{eqnarray}
\sum_j&\rightarrow& \frac{1}{s_0}\int d\vec{r}\nonumber\\
c_{ljvs}&\rightarrow& \sqrt{s_0}c_{lvs}(r) \nonumber
\end{eqnarray}
where $s_0 = \sqrt{3}/2$ is the area of the unit cell. Define the two component operator
\begin{equation}
C_{vs}(r) = 
\left[\begin{array}{c}c_{avs}(r)\\
c_{bvs}(r)
\end{array}\right],
\label{op1}
\end{equation}
and the $2\times 2$ matrix $\Sigma^{vs}(r,r')$ with the element given by
\begin{eqnarray}
\Sigma^{vs}_{ll'}(r,r') &=& (v_c\rho_l-sUm_l)\delta_{ll'}\delta(\vec r-\vec r')\nonumber\\
&& +v^{sc}(|\vec r-\vec r'|)\langle[c_{lvs}(r), c^{\dagger}_{l'vs}(r')]\rangle /2. \label{se}
\end{eqnarray}
We then get the MFT Hamiltonian in real space as
\begin{eqnarray}
H&=&\sum_{vs}[\int d\vec{r}C^{\dagger}_{vs}(r)h_{v}(\vec p)C_{vs}(r) \nonumber\\
&&+\int d\vec{r}\int d\vec{r'}C^{\dagger}_{vs}(r)\Sigma^{vs}(r,r')C_{vs}(r')], \label{mfh1}
\end{eqnarray}
the same as given by Eq. (\ref{mfh}).

Under a magnetic field $B$ perpendicular to the graphene plane, we work in the Landau level picture. 
With the wavefunction defined in Eq. (\ref{wv}), we have
\begin{eqnarray}
h_{v}(\vec p+\vec A)\phi^v_{nk}(r) = \phi^v_{nk}(r)\sqrt{2Bn}\sigma_1, 
\end{eqnarray}
and \cite{Ando}
\begin{eqnarray}
\int^{\infty}_{-\infty} dx \exp(iq_xx)\phi_{n}(x+k/B)\phi_{n'}(x+k'/B) \nonumber\\
= \sqrt{\frac{n_2!}{n_1!}}\exp(-\xi/2)\xi^{m/2}L^m_{n_2}(\xi)\exp(i\alpha) \nonumber
\end{eqnarray}
where $\xi=q^2/2B, n_1 = \max(n,n'), n_2 = \min(n,n'), m = n_1-n_2, L^m_{n_2}(\xi)$ is the Laguerre polynomial, and $\alpha = (n'-n)(\theta-\pi/2)-q_x(k+k')/2B$ with $\theta$ the angle between $\vec q$ and the $x$-axis. With these results, we get the formula for the MFT Hamiltonian in the LL picture as
\begin{equation}
H=\sum_{\lambda nkvs}\psi^{vs\dagger}_{\lambda}(n)[\sqrt{2Bn}\sigma_1+\Sigma^{vs}(n)]\psi^{vs}_{\lambda}(n)\hat a^{vs\dagger}_{\lambda nk}\hat a^{vs}_{\lambda nk} \nonumber
\end{equation}
with the self-energy matrix $\Sigma^{vs}(n)$ given by Eq. (\ref{sf}).

\section{Interaction matrix} 
\renewcommand{\theequation}{\thesection\arabic{equation}}
\setcounter{equation}{0}

Here we discuss how to calculate the elements of interaction matrix $v^K(n,n')$. The calculation algorithm is given for two cases of $\nu = 0$ and $\nu \ne 0$.
 
(A) $\nu = 0$. In this case, $v^{sc}(q) = v(q) = 2\pi e^2/q - 2\pi e^2/\sqrt{q^2+q^2_0}$. The integral in Eq. (\ref{vkn}) with the part $2\pi e^2/q$ in $v^{sc}(q)$ can be obtained analytically \cite{Gorbar} by using the formula 2.19.14.15 of Ref. \onlinecite{Prudnikov},
\begin{eqnarray}
&&\frac{1}{\pi}\int^{\infty}_0 dx \exp(-x)x^{m-1/2}L^m_j(x)L^m_{j'}(x) \nonumber\\
&=&\sum_{k=0}^{\min(j,j')}\frac{(-1)^{j+j'}\Gamma(1/2+k+m)}{k!(j-k)!(j'-k)!\Gamma(1/2+k-j)\Gamma(1/2+k-j')},\nonumber
\end{eqnarray}
where $j$ and $j'$ are integers. For the second part $2\pi e^2/\sqrt{q^2+q^2_0} \equiv v_s(q)$ (which is a short-range potential), there is no analytical formula available for the integral. Denote the corresponding matrix as $w(n,n')$ in the Landau picture [defined similarly by Eq. (\ref{vkn})]. In the large $\min(n,n')$ or large $|n-n'|$ limit, we have the semiclassical formula
\begin{eqnarray}
w_{ll}(n,n')&\to& \frac{B}{2\pi}\int_0^{\pi}\frac{d\theta}{\pi}v_s(|\vec k_{nl}-\vec k_{n'l}|), ~~~~l = 1,2, \nonumber\\
w_{12}(n,n')&\to& \frac{B}{2\pi}\int_0^{\pi}\frac{d\theta}{\pi}v_s(|\vec k_{n2}-\vec k_{n'2}|)\cos\theta, \nonumber
\end{eqnarray}
where $k_{n1} = \sqrt{2B(n-1)}, k_{n2} = \sqrt{2Bn}$, and $\theta$ is the angle between $\vec k_{nl}$ and $\vec k_{n'l}$. The azimuthal integrals in the above formula mean the angle average of the interaction of two electrons with momenta $k_{nl}$ and $k_{n'l}$. The factor $B/2\pi$ comes from the degeneracy of the state of momentum $k_{n'l}$.

(B) $\nu \ne 0$. In this case, since $\delta = s_0B\nu/4\pi$ and $q_{TF} \ne 0, v^{sc}(q)$ is a short-range potential. For large $\min(n,n')$ or large $|n-n'|$, we can use the above semi-classical approximation for $v^K(n,n')$. 

For $\min(n,n') > 50$ or $|n-n'| > 50$, the semi-classical approximation gives very accurate results with relative errors as the order of $10^{-3}$ or less. 

\section{high efficient approximation of series sum} 
\renewcommand{\theequation}{\thesection\arabic{equation}}
\setcounter{equation}{0}

The self-energy $\Sigma^{vs}_{ll'}(n)$ expressed in Eq. (\ref{sf}) is calculated by the sum over $n'$. The sum needs to be performed again and again for all $n$. At low magnetic field, the cutoff of $n'$ is very large. Meanwhile, one needs a large memory volume for storing the matrix function $v^v(n,n')$. To resolve the numerical difficulty, here we present a highly efficient scheme. 

\subsection{Approximate calculation of a sum}

First, we consider the following sum
\begin{equation}
S = \sum_{n=n_a}^{n_b}f(n).    \label{sum}
\end{equation}
where $n_a$ and $n_b$ are two integers for the lower and upper bounds of the sum. We suppose $f(x)$ as a function of the continuum variable $x$ is at least piecewise continuous and smooth. Then, we take parabolic approximation for $f(x)$ in small subintervals where the function is smooth. With the approximation, the summation can be carried out analytically. 

Suppose the function $f(x)$ is smooth in the interval $[n_1,n_3]$, we approximate it as
\begin{eqnarray}
f(x) \approx f(n_1) + c_1(x-n_1) + c_2(x-n_1)^2, \nonumber
\end{eqnarray}
where $c_1$ and $c_2$ are two constants. These constants can be determined with the knowledge of the function values $f(n_2)$ and $f(n_3)$ at two other points $n_2$ and $n_3$ with $n_1 < n_2 < n_3$,
\begin{eqnarray}
c_1 &=& \frac{n_3-n_1}{n_3-n_2}\frac{f(n_2) -f(n_1)}{n_2-n_1}-\frac{n_2-n_1}{n_3-n_2}\frac{f(n_3) -f(n_1)}{n_3-n_1}
, \nonumber\\
c_2 &=& \frac{f(n_3)-f(n_1)}{(n_3-n_1)(n_3-n_2)}-\frac{f(n_2)-f(n_1)}{(n_2-n_1)(n_3-n_2)}. \label{c2}
\end{eqnarray}
At the integer number $n$, we have
\begin{eqnarray}
f(n) \approx f(n_1) + c_1(n-n_1) + c_2(n-n_1)^2. \label{fj}
\end{eqnarray}
We hereafter suppose the three numbers $n_1, n_2$, and $n_3$ all are integers. Equation (\ref{fj}) is exact when $n$ equals any one of these three integers.

We now consider the following sum over the small interval $[n_1,n_3-1]$, 
\begin{eqnarray}
F(n_1,n_3) = \sum_{n=n_1}^{n_3-1}f(n). \label{s1}
\end{eqnarray}
Using the expressions (\ref{fj}) and (\ref{c2}), and
\begin{eqnarray}
\sum_{k=1}^{n}k &=& n(n+1)/2, \label{sm1}\\
\sum_{k=1}^{n}k^2 &=& n(n+1)(2n+1)/6, \label{sm2}
\end{eqnarray}
we get
\begin{eqnarray}
F(n_1,n_3) &\approx & w_1(n_1,n_2,n_3)f(n_1)+w_2(n_1,n_2,n_3)f(n_2)\nonumber\\
  &&+w_3(n_1,n_2,n_3)f(n_3) \nonumber
\end{eqnarray}
where the weight functions $w_{1,2,3}(n_1,n_2,n_3)$ are given by
\begin{eqnarray}
w_1(n_1,n_2,n_3) &=&\frac{n_3-n_1+1}{6(n_2-n_1)}(3n_2-2n_1-n_3+1), \nonumber\\
w_2(n_1,n_2,n_3) &=&\frac{(n_3-n_1)[(n_3-n_1)^2-1]}{6(n_2-n_1)(n_3-n_2)}, \nonumber\\
w_3(n_1,n_2,n_3) &=&\frac{n_3-n_1-1}{6(n_3-n_2)}(n_1-3n_2+2n_3-1). \nonumber
\end{eqnarray}

We now go back to the sum defined by Eq. (\ref{sum}). By selecting a sequence of an odd number of integers, $n_a = n_1 < n_2 < \cdots < n_{2m+1} = n_b$, we apply the above rule and get
\begin{eqnarray}
S &=& \sum_{\ell=1}^{m}F(n_{2\ell-1},n_{2\ell+1})+f(n_b) \nonumber\\
&=& \sum_{j=1}^{2m+1}W_jf(n_j) \label{fsm}
\end{eqnarray}
with
\begin{eqnarray}
W_1 &=& w_1(n_1,n_2,n_3), \nonumber\\
W_{2\ell-1}&=& w_1(n_{2\ell-1},n_{2\ell},n_{2\ell+1})\nonumber\\
&&~~+w_3(n_{2\ell-3},n_{2\ell-2},n_{2\ell-1}),\nonumber\\
&& ~~~~~~~~~~~~~~~~~~\ell = 2,3,\dots,m,\nonumber\\
W_{2\ell}&=& w_2(n_{2\ell-1},n_{2\ell},n_{2\ell+1}),
~~~~\ell = 1,3,\dots,m, \nonumber\\
W_{2m+1} &=& w_3(n_{2m-1},n_{2m},n_{2m+1})+1, \nonumber
\end{eqnarray}

In the present scheme, the selected integers are in ascending order, but not required to necessarily be equispaced. The choice of the properly distributed integers depends on the behavior of the function under sum. 

To test the accuracy and the efficacy of the scheme, we compare the numerical computations and the exact results for two examples below. 

{\it Example} 1. We consider the sum of the typical series 
\begin{eqnarray}
\zeta(p) &=& \sum_{n=1}^{\infty}\frac{1}{n^p},    \label{rmzf}
\end{eqnarray}
which is known as the Riemann zeta function. By numerical summation, $\zeta(p)$ is calculated by summing the terms up to a cutoff $N_c$. The error due to the dropped terms is about $O(N_c^{1-p})$. To suppress this error less than a small quantity $\delta$, we must have $N_c \sim \delta^{1/(1-p)}$. For $\delta = 10^{-4}$ and $p = 1.5$, one needs to sum $N_c = 10^8$ terms. Here, we select $M = 151$ integers $n_j$ as
\begin{eqnarray}
n_j = 
\begin{cases} [q^{j-1}], &{\rm if~} [q^{j-1}] > j \\
j, &{\rm otherwise} 
\end{cases}
\end{eqnarray}
where $q = 1.15$ and the square brackets mean the integer part of the number. Using the above numerical scheme, the sum is calculated by
\begin{eqnarray}
S = \sum_{j=1}^{M}W_j/n^p_j. \label{ssum}
\end{eqnarray}
The numerical results of this sum and the precise values of $\zeta(p)$ for various $p$ are listed in Table I. The error $\Delta=S-\zeta(p)$, stemming partly from the present scheme and partly from neglecting the terms $n > N_c$, seems satisfactorily small. Since the convergence of the sum is worse for smaller $p$, larger cutoff is needed for a high accuracy result. In the calculation, since the cutoff $[q^{M-1}]$ is the same for all parameters, the error is therefore larger for smaller $p$. 

\begin{table}[t]
\caption{\label{tab}Numerical sum $S$ compared with the Riemann function $\zeta(p)$ for various $p$. The quantity $\Delta=S-\zeta(p)$ represents the numerical error.}
\vspace{-3mm}
\begin{center}
\begin{ruledtabular}
\begin{tabular}{cccc}
$p$  & $\zeta(p)$ &$S$ & $\Delta$ \\
\hline
1.4 & 3.1055 &3.1048 &-0.0007 \\
1.5 & 2.6124 & 2.6122 &-0.0002\\
1.6 &2.2858& 2.2857  &-0.0001\\
1.8 & 1.8822 & 1.8822 &~0.0000\\
2 & 1.6449 & 1.6449 &~0.0000\\
\end{tabular}
\end{ruledtabular}
\end{center}
\end{table}

We define the efficacy $c$ as the ratio between the total number of terms to be summed in question and the actual number of terms in the approximate calculation. For the infinite series, the total number of terms is the cutoff number determined by the error tolerance $\delta$ from the dropped terms. Since the cutoff is $[q^{M-1}]$ in this calculation, the efficacy is
\begin{eqnarray}
c = [q^{M-1}]/M \approx 8.4\times 10^6.
\end{eqnarray}

{\it Example} 2. We numerically calculate the function $F(x)$ defined as,
\begin{eqnarray}
F(x) = \sum_{n=1}^{\infty}\frac{2x}{(n\pi)^2+x^2}. \label{ex2}
\end{eqnarray}
By setting the cutoff for the summation as $N_c = {\rm max}(10^5,10^5x/\pi)$, the error stemming from the dropped terms of $n > N_c$ is about $2x/N_c\pi^2$. For selecting the integers over which the sum is carried out, we note that $f(n)$ is flat for $n \ll x/\pi$. We therefore choose $m_0$ equispaced integers in the range $[1,N_0]$ with $N_0 = [4x/\pi]+1$. The total number of the summation is given as $M = 151$. The integer $m_0$ is set as $m_0=[4M/5]$ for $N_0 > [4M/5]$ or $m_0 = N_0$ for $N_0 \le [4M/5]$. The $M-m_0$ integers in the interval $(N_0, N_c)$ are chosen as
\begin{eqnarray}
n_j = 
\begin{cases} n_{j-1}+1, &{\rm for~} j=m_0+1,\cdots,j_0-1 \\
[qn_{j-1}], &{\rm for ~} j = j_0,\cdots,M
\end{cases}
\end{eqnarray}
where $q = (N_c/N_0)^{1/(M-m_0)}$ and $j_0$ is the minimum integer that $[qn_{j_0-1}] > n_{j_0-1}$. Thus, we have determined all the integers for the summation. With the selected integers, the weights can be calculated accordingly. By using the present sum method, we numerically calculate the function $F(x)$. The result for $F(x)$ is shown as the symbols in Fig. 6 and compared with the exact function (solid line) given by 
\begin{eqnarray}
F(x) = \tanh^{-1} x-x^{-1}. \nonumber
\end{eqnarray}
As seen from Fig. 6, the numerical results are surprisingly good. 

\begin{figure}[t] 
\centerline{\epsfig{file=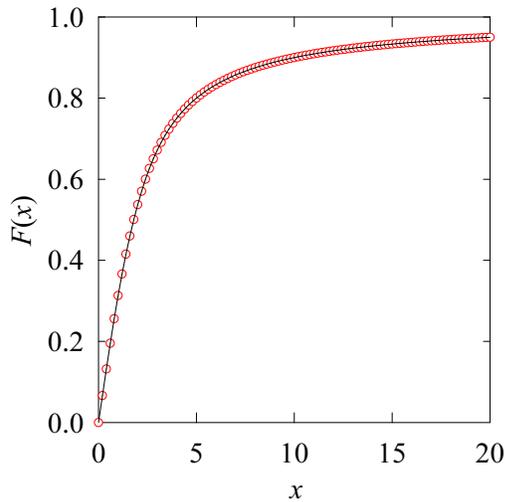,height=7.cm}}
\caption{(color online) Numerical results (symbols) for the function $F(x)$ compared with the exact formula (solid line).} 
\end{figure} 

\subsection{Calculation of the exchange self-energy}

We now apply the above numerical method to calculate the exchange self-energy in Eq. (\ref{sf}). For brevity, here we express the sum simply as
\begin{equation}
\Sigma^{xc}(n)=-\sum_{n'}v(n,n')g(n'). \label{xcse}
\end{equation}
The factor $g(n)$ corresponding to the distribution function is smooth except at the Fermi level. For describing $\Sigma^{xc}(n)$ and $g(n)$, we choose $M = 75$ points as  
\begin{eqnarray}
n_j = 
\begin{cases} [q^{j-1}], &{\rm if~} [q^{j-1}] > j \\
j-1, &{\rm otherwise} 
\end{cases}  \label{sn}
\end{eqnarray}
where $q = N_c^{1/(M-1)}$ and $N_c$ is the index of the largest Landau levels. The small $n_j$'s are actually adjacent integers (AIs). By taking $M = 75$ in the present calculation, for the magnetic field $B = 0.5$ T, the AIs range from 0 to 26. The Fermi-level index $\leq 3$ falls well into this range. In general, the AI range should be wide enough ensuring that there should be at least four AIs above the Fermi-level index $n_F$. Since the scheme presented in the previous section [or a quadratic polynomial interpolation that is used in the sum scheme and for interpolation of $g(n')$ again] assumes the function is local smooth, by so choosing the integers, the sum near the discontinuous point of $g(n)$ runs automatically over the AIs instead of using the approximation. 

On the other hand, as seen from Fig. 2 in the main text, the interaction $v(n,n')$ is not smooth at $n' = n$. Three cases need to be considered as follows. (i) When $n$ is very small so it falls deep (close to 0) into the range of AIs, the sum in Eq. (\ref{xcse}) can be performed using the above method with the points selected by Eq. (\ref{sn}) for $n'$. (ii) But, when $n$ goes out or from below close to the upper bound of the AI range, we need to choose a different set of $n'$. The selected numbers $n'$ should be densely distributed close to 0 and around $n$ making sure that besides an AI range $[0,n_F+k_1]$ with $k_1 \geq 4$ [for enclosing the discontinuous point of $g(n')$] there is another AI range $[n-k_2,n+k_3]$ with $k_2 $ and $k_3\geq 3$ [enclosing the spike of $v(n,n')$]. (iii) For $n = N_c$, the second AI range is $[N_c-k_2,N_c]$. The reason for doing these has been stated above since the present sum scheme or a quadratic polynomial interpolation requires the function to be local smooth. Therefore, the selection of $n'$ and thereby the weight for the sum depend on the point $n$. In the present calculation, the total number of points $n'$ for each $n$ is the same $M = 75$. The values of $g$ at the selected points $n'$ in the sum in Eq. (\ref{xcse}) for a given $n$ are obtained by interpolation from that at those selected points $n$. 
 
Shown in Fig. 3 of Sec. III are the self-energy $\Sigma^{K\uparrow}_{ll'}(n)$'s at $B = 0.5$ T and $\nu$ = 0. These functions vary smoothly with $n$. This fact shows that our selection of the integer points is reasonable and the calculation is reliable.

\section{Quantum Hall conductivity} 
\renewcommand{\theequation}{\thesection\arabic{equation}}
\setcounter{equation}{0}

The quantum Hall conductivity can be derived from the Kubo formalism. For simplicity, here we give a semi-classical derivation. By the classical theory, under a magnetic field $B$ in the $z$ direction, an electron moving in the $y$ direction with velocity $v$ feels a Lorentz force along the $-x$ direction
\begin{eqnarray}
F_x = -vB  \label{LF}
\end{eqnarray}  
in units of $e = c = \hbar =1$. To maintain the electron moving along the $y$ direction, we need to apply an electric field $E_x$ in the $-x$ direction
\begin{eqnarray}
E_x = -vB \equiv BJ_y/\rho_e \equiv J_y/\sigma_{yx}, \label{Ex}
\end{eqnarray}
where $J_y = -\rho_e v$ is the current density with $\rho_e$ as the electron density, and the Hall conductivity $\sigma_{yx}$ is so defined as
\begin{eqnarray}
\sigma_{yx}=\rho_e/B.
\end{eqnarray}

For electrons in graphene under a magnetic field, the (doped) electron density is calculated as (using the same notations as in Sec. II)
\begin{eqnarray}
\rho_e &=& \frac{1}{s_0}\sum_{l}(\langle n_l\rangle-1) \nonumber\\
  &=& \frac{1}{2s_0}\sum_{ls}\langle [c^{\dagger}_{ls}c_{ls}-c_{ls}c^{\dagger}_{ls}]\rangle \nonumber\\
  &=& \sum_{svkn\lambda}[f^{vs}_{\lambda}(n)-1/2]\psi^{vs\dagger}_{\lambda}\phi^{v\dagger}_{nk}(r)\phi^v_{nk}(r)\psi^{vs}_{\lambda}, \nonumber
\end{eqnarray}  
where the $l$-sum runs over the two sublattice indexes, and $n_l$ is the electron number on site $l$ in the unit cell. By the Landau gauge, the oscillator center is given as $x_c = -k/B$. With the help of relation $dk = Bdx_c$, after performing the $k$ sum, we obtain 
\begin{eqnarray}
\rho_e &=& \frac{B}{2\pi}\sum_{svn\lambda}[f^{vs}_{\lambda}(n)-1/2]\psi^{vs\dagger}_{\lambda}\psi^{vs}_{\lambda}  \nonumber\\
&=& \frac{B}{2\pi}\sum_{svn\lambda}[f^{vs}_{\lambda}(n)-1/2].  \nonumber
\end{eqnarray}  
Therefore, the Hall conductivity is given by
\begin{eqnarray}
\sigma_{yx} &=& \frac{1}{2\pi}\sum_{svn\lambda}[f^{vs}_{\lambda}(n)-1/2] ~~~~({\rm in~unit}~e^2/\hbar)\nonumber\\
&=& \sum_{svn\lambda}[f^{vs}_{\lambda}(n)-1/2] ~~~~({\rm in~unit}~e^2/h)\nonumber\\
&\equiv& \nu e^2/h \nonumber
\end{eqnarray}  
At CNP and at zero temperature, all the levels below (above) zero energy are fully filled (empty) while half of the zero levels are filled, which gives $\nu = 0$.

\end{document}